\documentstyle[aps,prl,multicol,epsf]{revtex}

\begin{document} 

\title{Density of states in the non-hermitian Lloyd model}

\author{Christopher Mudry, P. W. Brouwer, B. I. Halperin}
\address{Lyman Laboratory of Physics, Harvard University, Cambridge MA 02138}
\author{V. Gurarie, A. Zee}
\address{Institute for Theoretical Physics, University of California,
Santa Barbara CA 93106-4030}

\date{\today}
\draft

\maketitle 
 
\begin{abstract}
We reconsider 
the recently proposed connection between density of states in the
so-called ``non-hermitian quantum mechanics'' and the localization length
for a particle moving in random potential. We argue that it is indeed
possible to find the localization length from the
density of states of a non-hermitian random ``Hamiltonian''.
However, finding the density of states of a non-hermitian
random ``Hamiltonian'' remains an open problem, contrary to previous
findings in the literature. 
\end{abstract} 


\begin{multicols}{2}  

\section{Introduction}
\label{sec: Introduction}

There are situations in physics in which observables can be obtained
from properties of non-hermitian operators. In this context, the recent
work of Hatano and Nelson \cite{Hatano96} on random ``Hamiltonians'' with
an imaginary vector potential has caused considerable interest in
so-called ``non-hermitian quantum mechanics''
\cite{Efetov97,Brouwer97,Feinberg97:1,Janik97:1,Brezin98,Goldsheid97,Nelson97,Mudry97,Hatano98,Gurarie98,Silvestrov98}.
Motivated by the study of the pinning of vortices by columnar defects
in a superconductor, attention has focused on two main questions:  What
is the spectrum of eigenvalues of a non-hermitian ``Hamiltonian'',
and are the corresponding eigenfunctions localized or extended in
space?

In the model introduced by Hatano and Nelson, particles are hopping on
a lattice with a non-hermitian dynamics governed by the ``Hamiltonian''
\cite{Hatano96}
\begin{mathletters} \label{eq:HHN}
\begin{eqnarray}
&&
{\cal H}_h = 
{\cal K}_h + 
\sum_{{\bf j}} w^{\ }_{{\bf j}} \psi^{\dagger}_{{\bf j}} \psi^{\ }_{{\bf j}}, \\
\label{cal H_h}
&&
{\cal K}_h = -{t\over2} \sum_{{\bf j},{\bf a}}
\left(
e^{{\bf h} \cdot {\bf a}}
\psi^{\dagger}_{{\bf j}}\psi^{\ }_{{\bf j} + {\bf a}}+
e^{-{\bf h} \cdot {\bf a}}
\psi^{\dagger}_{{\bf j} + {\bf a}}\psi^{\ }_{{\bf j}}
\right).
\label{cal K_h}
\end{eqnarray}
\end{mathletters}%
Here, $\psi^{\dagger}_{\bf j}$ creates the state at lattice site ${\bf
j}$, ${\bf a}$ is a directed nearest-neighbor vector, $t$ is the
bandwidth, and $w_{{\bf j}}$ is the random (real) on-site potential.
Periodic boundary conditions are assumed.  The ``time evolution''
induced by ${\cal H}_h$ is non-unitary because of the imaginary vector
potential $i{\bf h}$.  
Numerical simulations by Hatano and Nelson support their conjecture 
that in the thermodynamic limit, the spectrum of the non-hermitian
operator (\ref{eq:HHN}) is concentrated on the real axis for energies
$\mbox{Re}\, \varepsilon \ge \varepsilon_0$ and extends into the
complex plane near the center of the band, 
$- \varepsilon_0 < \mbox{Re}\, \varepsilon < \varepsilon_0$. 
They showed that the eigenstates with real eigenvalues in the region
$|\mbox{Re}\, \varepsilon| > \varepsilon_0$ are localized, 
while the eigenstates corresponding to complex eigenvalues
are extended in space \cite{Hatano96}. 
The picture that emerges from their analysis is that
the energy $\varepsilon_0$, which separates the real and complex eigenvalues,
serves as a ``mobility edge'' for the non-hermitian problem
(barring some unforseen ``conspiracy'' in which all extended eigenfunctions 
in some energy range 
$\varepsilon_0-\epsilon<|{\rm Re}\, \varepsilon|<\varepsilon_0$
have real energy eigenvalues).
As was shown by Hatano and Nelson,
the value of the imaginary vector potential
${\bf h}$ where the eigenvalues of the non-hermitian ``Hamiltonian''
start to pop out into the complex plane is related to the localization
length $\xi(\varepsilon)$ of the problem in the absence of the
imaginary vector potential,
\begin{equation}
 |{\bf h}| = \xi(\varepsilon_0)^{-1} \label{eq:rel}.
\end{equation}

It was recently suggested by Hatano \cite{Hatano98} and Gurarie and Zee
\cite{Gurarie98} that the relationship (\ref{eq:rel}) can be inverted,
to use it as a method to extract the localization length of the
hermitian problem from the support of the spectrum of the non-hermitian
problem. In this way, knowledge of the support of the Density of States
(DoS) of the random Hamiltonian (\ref{eq:HHN}) as a function of the
imaginary vector potential permits the calculation of the localization
length as a function of energy in the hermitian case $h\equiv|{\bf h}|= 0$.

To our knowledge, the (ensemble averaged) non-hermitian DoS for the
random operator (\ref{eq:HHN}) is known only in zero \cite{Efetov97}
and one dimension \cite{Brouwer97,Goldsheid97}.  In this paper we
consider the DoS and its relation to the localization
length for the so-called Lloyd model \cite{Lloyd69}, in which the random potentials
$w_{\bf j}$ are independently distributed with the Cauchy distribution
\begin{equation} \label{eq:Cauchy}
  P(w) = {\gamma\over\pi}{1\over\gamma^2+w^2}.
\end{equation}
It is believed that the choice of the Cauchy distribution
(\ref{eq:Cauchy}) (instead of, say, a Gaussian one) does not modify the
universal properties of Anderson metal-insulator transition in dimension
$d>2$ \cite{Edwards 1971}.

As shown by Lloyd \cite{Lloyd69}, the advantage of this choice of the
probability distribution is that the ensemble averaged DoS of the
hermitian problem can be found exactly in any dimension $d$.  It has
been proposed in the literature that the DoS can also be obtained for
arbitrary $d$ in the case of the {\em non-hermitian} Lloyd model
\cite{Brezin98}. According to the relation (\ref{eq:rel}), such a
result would permit us to find the localization length of the hermitian
Lloyd model for arbitrary $d$. We show in this paper that the
calculation in \cite{Brezin98} does not give the correct DoS when
applied to dimensions $d>1$. We also discuss the difference between the
Lloyd model in $d=1$ and $d>1$ and illustrate for the one-dimensional
Lloyd model how the localization length for the hermitian problem can
be extracted from the support of the non-hermitian DoS. Calculation of
the localization length in the Lloyd model for $d \ge 2$, however,
remains an open problem.

\section{ Localization length from non-hermitian DoS }
\label{sec: Localization length from non-hermitian DoS }

We first discuss how one arrives at the relation (\ref{eq:rel}) between
the mobility edge $\varepsilon_0$, the imaginary vector potential ${\bf h}$,
and the localization length $\xi(\varepsilon)$ of the hermitian problem.

Hereto we consider, for a given realization of the random potential
$w_{\bf j}$, a non-degenerate eigenvalue $\varepsilon$ of the hermitian
operator ${\cal H}_0$ with periodic boundary conditions. 
(The subscript $0$ indicates that the imaginary vector potential $h$ is
set to zero, i.e., that the Hamiltonian ${\cal H}_0$ is hermitian.)
Following Ref.\ \onlinecite{Hatano96}, we assume that the corresponding
eigenstate $\Psi_0({\bf j})$ is localized by the random potential
$w_{\bf j}$, i.e., $\Phi_0({\bf j})$ is maximum at a site ${\bf m}$ and
decays exponentially far away from ${\bf m}$:
\begin{equation} \label{eq:Phi}
\Phi_0({\bf j})\sim 
\exp\left[-{|{\bf j}-{\bf m}|\over\xi(\varepsilon)}\right].
\end{equation}
By definition, the exponential decay length $\xi(\varepsilon)$ in
Eq.\ (\ref{eq:Phi}) is the localization length. Let us now switch on an
imaginary vector potential $i {\bf h}$.  As long as ${\bf h}$ is
sufficiently small, the wave function
\begin{equation}
\Psi({\bf j})= e^{{\bf h}\cdot{\bf j}}\ \Phi_0({\bf j})
\label{Ansatz}
\end{equation}
is a very good approximation to the exact eigenfunction $\Phi_h({\bf j})$ of 
${\cal H}_h$ which adiabatically evolves from $\Phi_0({\bf j})$ as $|{\bf h}|$
is increased. Although $\Psi$ satisfies 
${\cal H}_h \Psi = \varepsilon \Psi$, it is not an exact eigenfunction, 
because it violates the periodic boundary conditions. 
The error that one makes is of order $\exp\{[|{\bf h}|-1/\xi(\varepsilon)]L\}$.

 Hence, as long as 
\begin{equation}
|{\bf h}|<{1\over\xi(\varepsilon)}
\end{equation}
the wave function $\Psi$ will be a good approximation, and its energy
$\varepsilon$ will remain real and unshifted (up to an exponentially small
correction, in principle).

When the magnitude of the imaginary vector potential is larger than the
inverse localization length $1/\xi(\varepsilon)$, the wavefunction
(\ref{eq:Phi}) will no longer be a good approximation. Both the
eigenvalue and eigenfunction undergo a qualitative change reflecting
the non-hermiticity of the ``Hamiltonian''.  Hence, in the limit of an
infinite system size, at $|{\bf h}| = 1/\xi(\varepsilon)$, a generic
eigenvalue $\varepsilon$ enters the complex plane with unit
probability, resulting in the relation (\ref{eq:rel}).

To justify the inversion of Eq.\ (\ref{eq:rel}) to find the
localization length $\xi_(\varepsilon)$ from the support of the
spectrum of ${\cal H}_h$, we note that for $|{\bf h}| \approx
1/\xi(\varepsilon)$ eigenfunctions are strongly sensitive to the
boundary conditions. This sensitivity to the boundary conditions causes
the phenomenon of level attraction \cite{Feinberg97:1} with complex
eigenvalues coalescing along curves in $d=1$, or in compact sets
in $d \ge 2$ as the system size increases. The support of the
DoS of ${\cal H}_h$ appears to be self-averaging in the thermodynamic
limit, i.e., subject to decreasing fluctuations as the system size
increases. Therefore, the mobility edge $\varepsilon_0$ is well-defined
for the non-hermitian problem.  It has thus been proposed in
\cite{Hatano98} and \cite{Gurarie98} to relate the real part,
$\varepsilon_0$, of the energy at which the first complex eigenvalue
appears in the spectrum of ${\cal H}_h$ to the localization length
defined by Eq.\ (\ref{eq:rel}).


\section{ Single particle Green function for the non-hermitian ``Hamiltonian''}
\label{sec: Single particle Green function for the non-hermitian Hamiltonian}

The advantage of the Cauchy distributed disorder is that it allows the
exact calculation of the (ensemble averaged) single particle Green function.
In this section we discuss whether a similar property exists for 
a non-hermitian system with Cauchy disorder.

The DoS of the non-hermitian ``Hamiltonian'' (\ref{eq:HHN})
is computed from the Green function or resolvent
\begin{equation}
  {G_h(z)} = {1\over N}{\rm Tr}\, {1 \over z - {\cal H}_h},
\end{equation}
where $N$ is the total number of lattice sites (periodic boundary
conditions are assumed). The DoS $\rho_h(z)$ 
in the complex plane reads \cite{Feinberg97:1}
\begin{equation}
\label{analyt}
\rho_h(z) = 
{1\over\pi}{\partial\over\partial z^*} {G_h(z)}.
\label{DoS}
\end{equation}
For a hermitian system, when $G_0(z)$ is analytic for $\mbox{Im}\, z
\neq 0$ Eq.\ (\ref{DoS}) reproduces the usual DoS concentrated on the
real axis.

\begin{mathletters}\label{eq:BZ}
Let us now consider the ensemble average of the Green function $G_0$
and the DoS $\rho_0$. Lloyd \cite{Lloyd69} has shown that
the average Green function $\langle G_0 \rangle$ of the hermitian
Hamiltonian ${\cal H}_0$ is related to the Green function $K_0$ of the
non-random Hamiltonian ${\cal K}_0$ [see Eq.\ (\ref{eq:HHN})],
\begin{eqnarray} 
  && \langle{G_0(z)}\rangle = 
\sum_{\pm} 
{K}_0(z \pm {\rm i}\gamma)\theta(\pm{\rm Im}\, z),\\
  && {K}_0(z) = {1\over N}{\rm Tr}\, {1 \over z - {\cal K}_0}.
\end{eqnarray}
\end{mathletters}
The angular brackets denote an average over the random disorder
potential $w_{{\bf j}}$, $\gamma$ is the width of the distribution of
$w_{{\bf j}}$ [see Eq.\ (\ref{eq:Cauchy})], and $\theta(x) = 1$ $(0)$
for $x > 0$ ($x < 0$). 
It follows that the average DoS can be expressed in terms of the
non-random operator ${\cal K}_0$  only:
\begin{equation}
\langle\rho_0 (z)\rangle = 
{ 1 \over 2 \pi i} 
\left[
{K}_0(z-{\rm i}\gamma)-{K}_0(z+{\rm i}\gamma)
\right]
{\delta({\rm Im}\, z)}.
\label{averaged BZ DoS}
\end{equation}

Does the Green function relation Eq.\ (\ref{eq:BZ}) also hold for the
Green function $G_h$ of the non-hermitian ``Hamiltonian'' ${\cal H}_h$? The
answer is positive, provided 
\begin{equation}
|{\rm Im}\,  z| > \lambda \label{eq:necessary},
\end{equation}
where $\lambda$ is the imaginary part of the eigenvalue of the
non-hermitian ``Hamiltonian'' ${\cal H}_h$ with the largest imaginary part,
\begin{eqnarray}
\label{nec}
\lambda &=& 
{\rm sup}\ \{{\rm max}_k
{\rm Im}\, \varepsilon_k \} \nonumber \\
  &=& {\rm max}_k 
{\rm Im}\, \varepsilon'_k \nonumber \\
  &=& t \sinh (|{\bf h}| |{\bf a}|) \ \ \mbox{for $N \gg 1$}.
\end{eqnarray}
Here $\varepsilon_k$ ($\varepsilon'_k$), $k=1,\ldots,N$
are the $N$ eigenvalues of ${\cal H}_h$ (${\cal K}_h$) for a given
realization of the disorder and the supremum in the first line is taken
with respect to all possible disorder realizations.

To see why this is so, we choose to express the Green function in terms
of a replicated bosonic path integral
\begin{equation}
\label{path}
G_h(z)= {1 \over N} \mbox{Tr}\, 
\int {\cal D}[\phi^{\ }_\alpha,\phi^{* }_\alpha] \,
\phi^{* }_1 \phi^{\ }_1 
e^{\pm i \int \phi^{* }_\alpha\left( z-H_h \right) \phi^{\ }_\alpha}. 
\end{equation}
Here, $\alpha$ is a replica index.  The sign in the exponent is fixed by the
condition that the path integral be convergent. It is $+$ if ${\rm
Im}\,  z > \lambda$ and $-$ if  ${\rm Im} \, z < -\lambda$. If neither
of these two inequalities holds, i.e., if $z$ lies inside the strip
$|{\rm Im}\,  z| < \lambda $, the path integral (\ref{path}) cannot be constructed.
Averaging Eq.\ (\ref{path}) over the random potential $w_{\bf j}$ is
easily done with the Cauchy distribution of Eq.\ (\ref{eq:Cauchy}) if
$|\mbox{Im}\, z| > \lambda$. In that case, the replicated integrand
satisfies the condition of applicability of Cauchy Theorem after
closing the contour of integration over $w$ either in the upper half
plane or lower half plane, depending on whether the sign $+$ or $-$ is
chosen in Eq.\ (\ref{path}).
It is thus found that, for $|{\rm Im}\, z|>\lambda$,
\begin{mathletters}\label{eq:BZ h}
\begin{eqnarray} 
  && \langle{G_h(z)}\rangle = 
\sum_{\pm} 
{K}_h(z\pm{\rm i}\gamma) \theta(\pm{\rm Im}\, z), 
\ \ |\mbox{Im}\, z| > \lambda, \\
  && {K}_h(z) = {1\over N}{\rm Tr}\, {1 \over z - {\cal K}_h}.
\end{eqnarray}
\end{mathletters}
As in the hermitian case, the right hand side is expressed solely in
terms of the non-random resolvent ${\cal K}_h(z)$.

Equation (\ref{eq:BZ h}) first appeared in Ref.\ \onlinecite{Brezin98}
but without the restriction $|\mbox{Im}\, z| > \lambda$. 
The authors of Ref. \cite{Brezin98} applied Eq.\ (\ref{eq:BZ h})
to the strip $|{\rm Im} \, z| < \lambda $ 
to obtain the non-hermitian DoS in the complex plane
\begin{eqnarray}
\langle\rho_h(z)\rangle&&=
\sigma_h(z+{\rm i}\gamma)\theta({\rm Im}\, z)+
\sigma_h(z-{\rm i}\gamma)\theta(-{\rm Im}\, z)
\nonumber\\
&& \mbox{} + { 1 \over 2 \pi i} 
\left[
{K}_h(z-{\rm i}\gamma)-{K}_h(z+{\rm i}\gamma)
\right]
{\delta({\rm Im}\, z)},
\label{averaged BZ DoS h}
\end{eqnarray}
where $\sigma_h(z) = \pi^{-1} \partial_{z^{*}} K_h(z)$ is the DoS
of the non-hermitian problem in the absence of disorder.
The DoS (\ref{averaged BZ DoS h}) corresponds to a non-hermitian DoS 
coalescing both on the real axis (second line) and on a compact
set in the complex plane (first line).

The analytical continuation of Eq.\ (\ref{eq:BZ h}) to the strip
$|\mbox{Im}\, z| < \lambda$ in order to find the DoS can
be problematic since the Green function $G_h$ is a nonanalytic function
of $z$ where the DoS is nonzero [compare with Eq.\ (\ref{analyt})].
It can only be justified in the thermodynamic limit in one dimension
where the non-hermitian spectrum collapses to a $1d$ curve. We return
to this case in the next section. In all other cases Eq.\ (\ref{averaged
BZ DoS h}) is incorrect.
To illustrate where it might lead to, we consider Eq.\ (\ref{averaged BZ
DoS h}) in the thermodynamic limit $N \to \infty$ and extract the
length scale $l(\varepsilon) = 1/h$ by locating the edge
$\varepsilon(h)$ at which the wings of the spectrum fork into the
complex energy plane.
Using the arguments leading to Eq.\ (\ref{eq:rel}), one would identify
$l(\varepsilon)$ with the localization length.  The dependence of $l$
on energy $\varepsilon$ and dimensionality $d$ is then given by
\begin{eqnarray}
&& 
\cosh\left[{1\over l(\varepsilon)}\right] =
\cases
{
{1\over2t}\left[A_0(\varepsilon)+A_2(\varepsilon)\right],&$|\varepsilon|>   
(d-1)t$,\cr {1 \over t}
{\sqrt{\gamma^2+t^2}},     &$|\varepsilon|\leq(d-1)t$,\cr
}
 \nonumber \\ \label{l(E)} &&
  A_n(\varepsilon) = \sqrt{(|\varepsilon|-(d-n)t)^2    +\gamma^2}.
\end{eqnarray}
Note that the length scale $l(\varepsilon)$ given by Eq.\ (\ref{l(E)})
is finite for all energies and all
dimensions. Hence, if Eq.\ (\ref{averaged BZ DoS h}) were true, one
would conclude that, irrespective of dimensionality, all states are
localized in the Lloyd model \cite{Gurarie98}.  This conclusion is not
surprising in $1d$ or $2d$.  In fact, the length $l(\varepsilon)$
agrees with the localization length of the Lloyd model in $1d$
\cite{Hirota71,Thouless72,Eggarter74}.  
In $2d$, however, $l(\varepsilon)$ is much smaller than the weak disorder
estimate for a Gaussian disorder \cite{Lee85}.  Moreover, in $d>2$,
such a conclusion contradicts the belief that the existence of large
tails in the Cauchy distribution does not modify the universal
properties of Anderson metal-insulator transition \cite{Edwards 1971}.
We return to the issue of dimensions $d\geq2$ and the interpretation of
$l(\varepsilon)$ in section \ref{sec: Higher dimensions: mean free path
}.  The reason why the length scale $l(\varepsilon)$ cannot be
interpreted as the localization length for $d > 1$ is that analytical
continuation of Eq.\ (\ref{eq:BZ h}) to the strip $\mbox{Im}\, z <
\lambda$ is in general invalid unless the DoS is supported on a one
dimensional curve, like in $1d$ or in the hermitian case ${\bf h}=0$.
In particular, we conclude that Eq.\ (\ref{averaged BZ DoS h}) does
not yield the average DoS of the non-hermitian extension of Lloyd
model in $d>1$.

\section{Non-hermitian DoS for one chain} 
\label{sec: Non-hermitian DoS for one chain }

In view of the unreliability of analytic continuation of Eq.
(\ref{eq:BZ h}), it is important to compare Eq.\ (\ref{eq:BZ h}) with
what is known about the spectral properties of the non-hermitian
``Hamiltonian'' ${\cal H}_h$ from other methods.

First, we note that in any dimension, analytic continuation of Eq.
(\ref{eq:BZ h}) is certainly wrong in a system of finite size: For any
finite system and for any dimension the support of the averaged DoS
$\langle\rho_h(z)\rangle$ occupies the entire strip in the complex
energy plane that is excluded in Eq.\ (\ref{eq:necessary}). To see this,
choose the realization $w_1=\cdots=w_N=V$ with $V$ an arbitrary real
number. Equation (\ref{eq:BZ h}), however, results in a DoS with a
significantly smaller support: it is the DoS of the system without
disorder shifted by an amount $\pm \gamma$ towards the real axis
\cite{Brezin98}.

What about the DoS in the thermodynamic limit $N \to\infty$? Let us
first discuss the one-dimensional Lloyd model. A discussion of the case
$d>1$ is postponed to the next section.  In one dimension, several
independent approaches have been taken in the literature
\cite{Brouwer97,Janik97:1,Goldsheid97}. For the Lloyd model, Goldsheid
and Khoruzhenko \cite{Goldsheid97} have shown that the support of the
spectrum of ${\cal H}_h$ is self-averaging in the thermodynamic limit and
found a DoS that coincides with Eq.\ (\ref{averaged BZ DoS h}).  Hence,
the DoS obtained from Eq.\ (\ref{averaged BZ DoS h}) is correct in an
infinite one-dimensional system, despite the flaws in its derivation.
Starting from this non-hermitian DoS, one can use the arguments of
section \ref{sec:  Localization length from non-hermitian DoS } to
identify $l(\varepsilon)$ with the localization length
$\xi(\varepsilon)$ of the Lloyd model in one dimension. 

We find it instructive to present an alternative derivation of Eq.
(\ref{averaged BZ DoS h}) for weak disorder, using the approach of
Ref.\ \onlinecite{Brouwer97}, where the support of the DoS was
calculated for weak Gaussian disorder.
In this approach, knowledge of the localization length is required
to calculate the non-hermitian DoS.
In the absence of disorder, the energy spectrum is parameterized in
terms of the (complex valued)
wavenumbers $s={2\pi/N}+ih,\cdots,2\pi+ih$ 
of the plane wave states diagonalizing
${\cal K}_h$,
\begin{equation}
\varepsilon'(s)=-t\cos s.
\end{equation}
In Ref.\ \onlinecite{Brouwer97}, a transfer-matrix approach was used to
calculate the spectrum of ${\cal H}_h$ for weak non-hermiticity and
weak disorder to leading order in $1/N$. Weak non-hermiticity means
$|{\bf h}||{\bf a}| \ll 1$, whereas weak disorder amounts to $ |\sin {\rm
Re}\, s|\, \xi \gg 1 $, where $\xi$ is the localization length of the
$1d$ hermitian Lloyd model, see Eq.\ (\ref{l(E)}).
To leading order in $1/N$, it was found that \cite{Brouwer97}
\begin{mathletters}
\label{transfer matrix methods}
\begin{eqnarray}
|{\rm Im}\, s|&&=
|h|-\xi^{-1}
\label{general properties of trsf matrix}\\
&&=
|h|-{\gamma\over\sqrt{t^2-({\rm Re}\, \varepsilon)^2}}+
{\cal O}({\gamma^2/t^2}).
\label{specificity of Cauchy}
\end{eqnarray}
\end{mathletters}
Here, we have expanded Eq.\ (\ref{l(E)})
to lowest non-trivial order in $\xi^{-1}$ and $\gamma/t$. Hence, we find
that the support of the spectrum is given by
\begin{eqnarray}
{\rm Im}\,  \varepsilon &=& -t \cos ({\rm Re}\, s) \sinh({\rm Im}\, s) 
  \nonumber \\ &=& 
  \pm h\sqrt{t^2-({{\rm Re}\, \varepsilon})^2}\mp\gamma,
\label{Piet's derivation of support in d=1}
\end{eqnarray}
in agreement with Eq.\ (\ref{averaged BZ DoS h}) when $d=1$. 
The variance of ${\rm Im}\, \varepsilon$ vanishes like $N^{-1}$ 
in the thermodynamic
limit. Note that Eq.\ (\ref{Piet's derivation of support in d=1}) 
relies on very general properties of
one-dimensional disordered systems through Eq.\ (\ref{general
properties of trsf matrix}) and only on the specificities of the
hermitian Lloyd model through Eq.\ (\ref{specificity of Cauchy}).
One dimension is very special in that localization length
and DoS are closely related \cite{Thouless72}.
The situation is more intricate in higher dimensions where a new length
scale, the mean free path, appears besides the localization length.

For small $\gamma$, the DoS in the Lloyd model is
qualitatively different from the non-hermitian DoS in the presence of
Gaussian distributed disorder \cite{Brouwer97,Goldsheid97}: In the
latter case, the uniform shift $\gamma$ in Eq.\ (\ref{Piet's derivation
of support in d=1}) should be replaced by the non-uniform and much
smaller shift $\gamma^2/\sqrt{t^2-({\rm Re}\, \varepsilon)^2}$, where
$\gamma^2$ is the variance of the Gaussian distribution.  To
explain the difference, we compare the $\gamma$-dependences of the
localization lengths at the center of the band for both disorder
distributions. For the Cauchy distribution we have $\xi(0)=t/\gamma$,
while $\xi(0)=t^2/\gamma^2$ for a Gaussian distribution
\cite{Dorokhov88}. Using the relation (\ref{transfer matrix
methods}) between localization length and non-hermitian DoS, such a
difference is directly carried over to the DoS.
In fact, the mechanism of localization in $1d$ is very different for
Cauchy disorder compared to that of Gaussian disorder: For a Cauchy
distribution, the dependence on the disorder strength $\gamma$ of the
inverse localization length $\xi^{-1}\sim\gamma/t$ follows immediately
from estimating the probability that the disorder potential $w_{\bf j}$
at an arbitrary site be larger than the band width $t$, in which case
the chain is classically broken. Thus, localization in the one
dimensional hermitian Lloyd model is not caused by quantum
interferences effects, but rather by wave functions accomodating to
large fluctuations in the disorder by vanishing locally. In contrast,
in the case of Gaussian distributed disorder, localization is entirely
due to quantum interference.

\section{Higher dimensions: mean free path}
\label{sec: Higher dimensions: mean free path }

In section \ref{sec: Single particle Green function for the
non-hermitian Hamiltonian} we have shown that analytical continuation
of Eq.\ (\ref{eq:BZ h}) into the strip $|\mbox{Im}\, z| \le \lambda$
yields a length scale $l(\varepsilon)$ that remains finite irrespective
of dimensionality. If this analytical continuation were justified, the
arguments of section \ref{sec: Localization length from non-hermitian
DoS } would allow us to identify $l(\varepsilon)$ as the localization
length of Lloyd model. Then the Lloyd model would not display a
metal-insulator transition irrespective of dimensionality. However, as
we have seen, in general, Eq.\ (\ref{eq:BZ h}) cannot be applied inside
the strip $|\mbox{Im}\, z| \le \lambda$. This does not mean that the
length scale $l(\varepsilon)$ obtained from Eq.\ (\ref{eq:BZ h}) is
entirely meaningless. In this section, we compare the length scale
$l(\varepsilon)$ defined by Eq.\ (\ref{l(E)}) with other length scales
that appeared in previous studies of the Lloyd model
\cite{Johnston83,Thouless83,MacKinnon84,Rodrigues86}, and find that, in
the limit of weak disorder, $l(\varepsilon)$ is to be interpreted as
the mean free path, rather than the localization length. [For strong
disorder there is no distinction between mean free path and
localization length.]
The fact that analytical continuation of Eq.\ (\ref{eq:BZ h}) yields
the correct DoS and localization length in one dimension is thus simply
understood as the fact that localization length and mean free path
coincide in $d=1$.

To see what is the correct interpretation of the length scale
$l(\varepsilon)$ defined in Eq.\ (\ref{l(E)}), 
we go back to the work of Johnston and Kunz \cite{Johnston83}, 
who considered the quantity
\begin{equation}
{1 \over l'_r(z)} = -{r}^{-1}
\left\langle
\ln
\left|
{\cal G}_{{j},{j}+{r}}(z)
\right|
\right\rangle,
\label{alt def 2}
\end{equation}
where ${\cal G}(z) = (z-{\cal H}_0)^{-1}$.
They found an expression for Eq. (\ref{alt def 2}) valid at least for
sufficiently large $|{\rm Im}\, z|$, at fixed $r$ and system size.
Johnston and Kunz also found that if their expression for $l'_r(z)$
is analytically continued to $z\to\varepsilon$ on the real axis,
and the limits of infinite system size and separation $r$ are then taken,
one obtains a result $l'(\varepsilon)$ which coincides with 
Eq. (\ref{l(E)}). Finally, 
Johnston and Kunz hypothesised that $l'(\varepsilon)$
should coincide with the localization length, which one may define by
\begin{equation}
{1\over \xi(\varepsilon)} = - \left\langle
\lim_{r \to \infty} {r}^{-1} \lim_{z \to \varepsilon}
\ln
\left|
{\cal G}_{{j},{j}+{r}}(z)
\right| \right\rangle.
\label{alt def 1}
\end{equation}
[We remark that the localization length $\xi(\varepsilon)$ is a
self-averaging quantity, so that the ensemble average in Eq.\ (\ref{alt
def 1}) is not necessary.]
However, as was noticed by
Thouless \cite{Thouless83},
MacKinnon \cite{MacKinnon84} and by 
Rodrigues, Pastawski, and Weisz \cite{Rodrigues86}, 
this hypothesis is not correct. Moreover, Rodrigues, Pastawski, and Weisz 
showed that the result (\ref{l(E)}) for $l'(\varepsilon)$
coincides with the length scale $l''(\varepsilon)$ defined by
\begin{equation}
{1\over l''(\varepsilon)}=
-\lim_{r \to \infty} {r}^{-1} 
\ln\left|\lim_{z \to \varepsilon}
\left\langle
{\cal G}_{{j},{j}+{r}}(z)
\right\rangle
\right|.
\label{kappa(varepsilon)}
\end{equation}
This length scale is naturally interpreted as the scale on which the phase
of wave functions is randomnized, rather than the length
scale on which the amplitude decays exponentially.
The length $l''(\varepsilon)$ can also be interpreted as the mean
free path of a particle for weak scattering. Unfortunately, to find the
true localization length $\xi(\varepsilon)$, defined by Eq. (\ref{alt def 1}),
remains an open problem in the Lloyd model in $d>1$.

A more intuitive picture of what is going on in higher dimensions can be
obtained along the lines of the last paragraph of section 
\ref{sec: Non-hermitian DoS for one chain }.
The length scale $l(\varepsilon)$ of Eq.\ (\ref{l(E)}) behaves like
\begin{equation} \label{eq:l(E)simp}
  l(\varepsilon) =
  {t\over\gamma}+{\cal O}\left({\gamma^2\over t^2}\right),\quad
|\varepsilon|<(d-1)t,
\end{equation}
for all dimensions and weak disorder. This implies that the calculation
scheme of Ref. \cite{Brezin98}, i.e., application of Eq.\ (\ref{eq:BZ h})
to the strip $|{\rm Im}\, z|<\lambda$,
is predicated on the same mechanism as in $1d$,
namely the removal of a site with probability $\gamma/t$. In contrast
to $1d$, the removal of a site does not preclude propagation in $d \ge2$.  
Instead, the removal of sites leads to the usual impurity
scattering, and we see from Eq.\ (\ref{eq:l(E)simp}) that
$l(\varepsilon)$ has the semiclassical
interpretation of an average length for free propagation between two
impurities, i.e., a semiclassical mean free path.


If we are to interpret Eq.\ (\ref{l(E)}) for weak disorder
as the mean free path, we have
no reason to exclude that one parameter scaling \cite{Lee85} applies to
${\cal H}_h$ in the case ${\bf h}=0$. It is widely
believed that propagation becomes diffusive for a
window of length scales beyond the mean free path $l(\varepsilon)$. The
upper length scale for diffusive propagation is by definition, the
localization length. Thus, for the Lloyd model in the weak disorder limit
we are lead to expect
the same localization properties as for weak Gaussian distributed disorder:
(i)  localization at all energies if $d=2$, (ii) existence of a
diffusive regime for a window of energies centered around $\varepsilon=0$ if
$d=3$. This expectation is confirmed by the localization properties
\cite{Raghavan81} of a caricature of Lloyd model defined by a random
hopping tight binding Hamiltonian whereby the hopping amplitude takes
the value $t$ with probability $1-\gamma/t$ and vanishes with
probability $\gamma/t$.

\section{Conclusion}
\label{sec: Conclusion}

In this paper we have outlined that the localization length of a
particle moving in a random potential can be obtained from the averaged
DoS of a non-hermitian particle in a random potential.
Non-hermitian DoS have only been calculated in zero 
and one dimensions.
The calculation of the non-hermitian DoS in closed form in more than
one dimension thus remains an open problem.
We expect that the non-hermitian trick should provide an
alternative to numerical calculations of the localization length
relying on transfer matrix approaches in strip geometries.

We are indebted to J. T. Chalker and D. R. Nelson for useful
discussions.  We acknowledge the support by the NSF under grants
no.\ DMR 94-16910, DMR 96-30064, DMR 97-14725, and PHY 94-07194
and from the Swiss Nationalfonds (CM).

\end{multicols}  
\end{document}